\documentclass{PoS}

\title{
  \begin{picture}(0,0)(0,0)%
    \put(350,75){\makebox(0,0)[l]{\textnormal{\normalsize YITP-15-99}}}%
    \end{picture}
 Lattice QCD studies on baryon interactions \\
 from L\"uscher's finite volume method
 \\ and 
 HAL QCD method 
 }
\ShortTitle{
 Lattice QCD studies on baryon interactions from 
 L\"uscher's finite volume method
 and 
 HAL QCD method 
}
 \author{\speaker{Takumi Iritani}$^{ab}$ \\
  $^a$ Department of Physics and Astronomy, Stony Brook University, \\
  Stony Brook, New York 11794-3800, USA \\
  $^b$ Yukawa Institute for Theoretial Physics (YITP), Kyoto 606-8502, Japan \\
E-mail: 
\email{takumi.iritani@stonybrook.edu},
\email{iritani@yukawa.kyoto-u.ac.jp}}

\author{for HAL QCD Collaboration\\
  \includegraphics[width=0.35\textwidth]{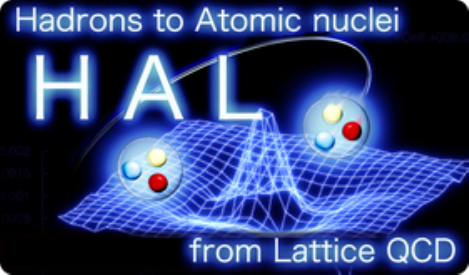}
}

\abstract{
A comparative study between the L\"uscher's finite volume method and the time-dependent 
HAL QCD method is given for the $\Xi\Xi$($^1\mathrm{S}_0$) interaction as an illustrative example.
By employing the smeared source and the wall source for the interpolating  operators, 
we show that the effective energy shifts $\Delta E_{\rm eff} (t)$ in L\"uscher's method
do not agree between different sources,
yet both exhibit fake plateaux.
  On the other hand, the interaction kernels $V(\vec{r})$ obtained from the two sources in the HAL QCD method
 agree with each other already for modest values of $t$.
 We show that the energy eigenvalues $\Delta E(L)$ in finite lattice volumes ($L^3$) 
  calculated by $V(\vec{r})$ indicate that 
   there is no bound state in the $\Xi\Xi(^1\mathrm{S}_0)$ channel
   at $m_{\pi}=0.51$ GeV  in 2+1 flavor QCD. 
}

\FullConference{The 33rd International Symposium on Lattice Field Theory\\
		14 -18 July 2015\\
		Kobe International Conference Center, Kobe, Japan*}

\begin{document}

\vspace{-2ex}
\section{Introduction} 
\vspace{-1ex}

To investigate the hadron-hadron interactions, two
methods have been proposed so far; the L\"uscher's finite volume method
\cite{Luscher:1991} and   the HAL QCD method \cite{Ishii:2006ec}.  In the
former method, the energy shift of the two-body system in finite lattice
box(es) is measured.  It is then translated into the scattering phase shift and
the binding energy in the infinite volume through the L\"uscher's formula (see
e.g. a review, \cite{FVM-review}).  On the other hand, in the latter method,
the interaction kernel (the non-local potential) between hadrons is first
calculated in finite lattice box(es).  It is then utilized to  calculate the
observables in the infinite volume (see the review, \cite{Aoki:2012tk}.)

For the simple example such as the $I = 2$ $\pi\pi$ scattering with the heavy
pion mass, a quantitative agreement of the phase shifts and
the scattering lengths between the two methods has been established
\cite{Kurth:2013tua}.  On the other hand, no systematic comparison between two
methods for multi-baryon systems has been made so far. The purpose of the present
report  is to make such a comparison in the two-baryon system 
on the common
gauge configurations.  As shown below, our results indicate that an extreme
care is necessary for the  L\"uscher's method when it is applied to
baryon-baryon interactions.

\vspace{-1ex}
\section{Lattice Setup}
\vspace{-1ex}

We use 2+1 flavor QCD gauge configurations generated with the Iwasaki gauge
action and the 
nonperturbatively ${\cal O}(a)$-improved 
Wilson quark action at $\beta = 1.90$ and $c_{sw} = 1.715$, which
corresponds to $a^{-1} = 2.194(10)$ GeV, $m_\pi = 0.51$ GeV,  $m_N = 1.32$ GeV,
and $m_\Xi = 1.46$ GeV 
\cite{Yamazaki:2012hi}.
We take the three
lattice volumes, $L^3\times T$ = $40^3 \times 48$, $48^3 \times 48$ and $64^3 \times 64$, which
correspond to the spatial sizes 3.6 fm, 4.3 fm and 5.8 fm, respectively.  These
configurations are {\it exactly} the same as those used by Yamazaki {\it et
al.} for nucleon-nucleon (NN) systems~\cite{Yamazaki:2012hi}.  In order to improve the statistics, we make a
use of the rotation symmetry for $48^3 \times 48$ and $64^3 \times 64$
lattices. 

In this report, we focus on the baryon-baryon interaction in the
$\Xi\Xi(^1\mathrm{S}_0)$ channel instead of the NN channel.
This is because 
the statistical error in the hyperon system is much smaller than those of the
nucleon system due to the strange quark mass,  so that the quantitative
comparison between the two methods can be made clearer.  Also, the
$\Xi\Xi(^1\mathrm{S}_0)$ channel belongs to the same SU(3) flavor multiplet (the {\bf
27} representation) as the NN$(^1\mathrm{S}_0)$,  so the characteristic
features of the interaction are expected to be similar.
We employ 
interpolating operators for $\Xi$ as 
\begin{equation}
  \Xi_0^\alpha = \varepsilon^{abc} (s_a C \gamma_5 u_b) s_c^\alpha,
  \quad
  \Xi_-^\alpha = \varepsilon^{abc} (s_a C \gamma_5 d_b) s_c^\alpha
  \label{eq:Xi_operator}
\end{equation}
with relativistic (4-spinor) quark fields and $C = \gamma_4 \gamma_2$.
The quark propagator is solved with the periodic boundary condition in all directions.

With Coulomb gauge fixing,  we employ both smeared source and wall source 
for the interpolating 
 operators to check whether observables are independent of the choice.  As for
 the wall source, we adopt 
 $q^{\mathrm{wall}}(t_0) = \sum_{\vec{x}} q(\vec{x},t_0)$, while for the
 smeared source, we take the exponentially smeared quark, 
 $q^{\mathrm{smear}}(\vec{x},t_0) = \sum_{\vec{y}} f(|\vec{x}-\vec{y}|) q(\vec{y},t_0)$
 where $f(r \neq 0) = A e^{-Br}$ and $f(r = 0) = 1$
with coefficients $A$, $B$ taken from \cite{Yamazaki:2012hi},
so that the smeared source is exactly the same as \cite{Yamazaki:2012hi}.
Our lattice parameters  are summarized in
 Table~\ref{tab:lattice_seups}.
 For the smeared source,
(\# conf. $\times$ \# sources) in Yamazaki et al. ~\cite{Yamazaki:2012hi} is
(200 $\times$ 192), (200 $\times$ 192) and (190 $\times$ 256)
for $L^3=40^3$, $48^3$ and $64^3$, respectively, and 
the ratio of the statistics in this work to Yamazaki et al. is about 1.0, 5.3 and 0.32
for each $L$.

\begin{table}[h]
  \centering
  \begin{tabular}{c|c|cc|c}
    \hline \hline
    volume & \# conf. & \# smeared source & parameter $(A,B)$ & \# wall source \\
    \hline
    $40^3 \times 48$ & $200$ & $192$          & $(0.8,0.22)$ & $48$ \\
    $48^3 \times 48$ & $200$ & $4 \times 256$ & $(0.8,0.23)$ & $4 \times 48$ \\
    $64^3 \times 64$ & $327$ & $1 \times 48$  & $(0.8,0.23)$ & $4 \times 32$ \\
    \hline \hline
  \end{tabular}
  \caption{
Lattice QCD setup.
  For some of the \# source, an extra factor 4 from rotations is presented.}
  \label{tab:lattice_seups}
\end{table}

\vspace{-3ex}
\section{L\"uscher's Finite Volume Method}

\vspace{-2ex}
\subsection{Effective energy shift}
\vspace{-1ex}

In the L\"uscher's method, the key quantity is the  
 energy shift of the two-body system in the finite volume, $  \Delta E(L) = E^{BB} - 2 m^B$,
where $E^{BB}$ is the ground state energy of the two baryons  and $m^B$ is the mass of a single baryon.
The L\"uscher's formula relates $k$ of  $\Delta E(L) = 2\sqrt{(m^B)^2 + k^2} - 2m^B$
to  the phase shift $\delta(k)$ in the infinite volume as \cite{Luscher:1991}
\begin{equation}
  k \cot \delta(k) = \frac{1}{\pi L} \sum_{\vec{n} \in \mathbf{Z}^3}
  \frac{1}{|\vec{n}|^2 - |\frac{\vec{k}L}{2\pi}|^2}.
\end{equation}

In practice,  $\Delta E(L)$ is often obtained 
from the plateau of the effective energy shift
$\Delta E_\mathrm{eff}(t)$ at large $t$~\cite{Yamazaki:2012hi},
\begin{equation}
  \Delta E_\mathrm{eff}(t) = E_\mathrm{eff}^{BB}(t) - 2 m_\mathrm{eff}^{B}(t),
  \label{eq:Eeff}
\end{equation}
where 
$m_\mathrm{eff}^{B}(t)=\ln (G^{B}(t)/G^{B}(t+1))$ and 
$E_\mathrm{eff}^{BB}(t)=\ln (G^{BB}(t)/G^{BB}(t+1))$
with
$G^{B}(t)$ and $G^{BB}(t)$ being the correlation functions of a single baryon
and two baryons, respectively. 
An advantage of taking the difference $\Delta E_\mathrm{eff}(t)$
is that a correlation between $G^{B}(t)$ and $G^{BB}(t)$ on each configuration 
makes the absolute magnitude of the statistical error 
for $\Delta E_\mathrm{eff}(t)$ much smaller than that for $G^{BB}(t)$.
In addition, 
the contaminations from single-baryon excited states at small $t$  in $\Delta E_\mathrm{eff}(t)$
is expected to be cancelled in part between $E_\mathrm{eff}^{BB}(t)$ and  $2m_\mathrm{eff}^{B}(t)$. 
On the other hand, the contaminations from the elastic $BB$ scattering states other than
the ground state appear only in $E_\mathrm{eff}^{BB}(t)$ and therefore they propagate into 
$\Delta E_\mathrm{eff}(t)$.  Furthermore, such contaminations  on a large lattice box 
 survive  even at large $t$.
  For example, the first excited state of the elastic $\Xi \Xi$ scattering has only
  $\sim 30$ MeV excitation energy in $L^3=64^3$ in the present setup:  Such an
  excited state can be isolated only for $t \gg (30 \ \mathrm{MeV})^{-1} \sim \mathcal{O}(10)$ fm.
  This consideration indicates that fitting a plateau-like structure of  
  $\Delta E_\mathrm{eff}(t)$  for moderate values of $t$ is 
  dangerous to extract the true signal in the L\"uscher's method.
 Indeed, in the next subsection,  we demonstrate explicitly that 
 such a fake plateau appears.

\vspace{-1ex}
\subsection{Source operator dependence}
\vspace{-1ex}

\begin{figure}
  \centering
  \includegraphics[width=0.47\textwidth,clip]{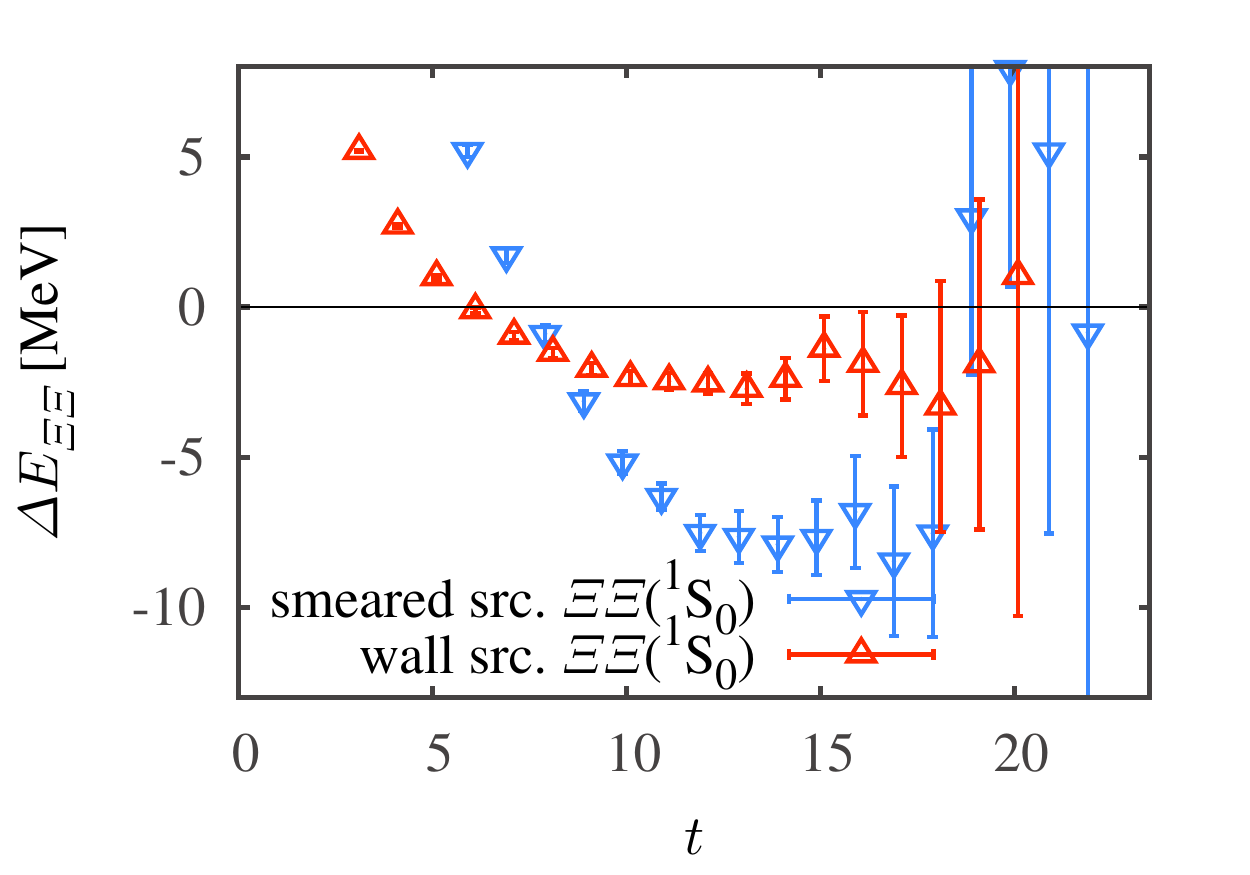}
  \caption{
    \label{fig:energy_shift}
    The effective energy shift $\Delta E_\mathrm{eff}(t)$
    using the smeared source (blue down triangle) and the wall source (red up triangle) on a $48^3 \times 48$ lattice.
    Both show plateau-like behaviors for $t=12-16$, whose values, however,
    are significantly different with each other.
  }
\end{figure}

One of the  useful methods to identify the fake plateau is to examine the
source operator dependence of the effective energy shift.
Shown in Figure \ref{fig:energy_shift}  
is $\Delta E_\mathrm{eff}(t)$ on a $48^3 \times 48$ lattice
for the smeared source (blue) and the wall source (red).
We find plateau-like structures for  both sources in the range $t = 12-16$. 
Their magnitudes, however,  do not agree with each other within statistical errors.
This casts a strong doubt on the validity of plateaux in  Figure \ref{fig:energy_shift}:
Either one  of the plateaux (or both) is fake 
 and the real plateau would appear for much larger $t$ where 
  higher statistics are required to extract 
a few to $\sim$ 10 MeV energy shift.
This analysis suggests that  rather strong binding
of the two-baryon systems, claimed in Refs. \cite{Yamazaki:2012hi,Yamazaki:2015asa}
by using $\Delta E_\mathrm{eff}(t)$, should be taken  with a grain of salt. 
The same caution applies also to the results of e.g. Ref.\cite{Beane:2011iw}.

\vspace{-1ex}
\section{HAL QCD Method}
\vspace{-2ex}

\subsection{Interaction kernel}
\vspace{-1ex}

In the  ``time-dependent" HAL QCD method \cite{HALQCD:2012aa},
 we consider the Nambu-Bethe-Salpeter (NBS) correlation function $R(\vec{r}, t)$; 
\begin{equation}
  R(\vec{r}, t) \equiv
  \left\langle 0 | T\{B(\vec{x} + \vec{r},t)
  B(\vec{x},t)\} \bar{\mathcal{J}}(0)| 0 \right\rangle
  /\left\{ G^B(t) \right\}^2
  = \sum_n A_n \phi^{W_n}(\vec{r}) e^{-\Delta W_n t}
  + \mathcal{O}(e^{-\Delta W_{\mathrm{th}} t}).
  \label{R_correlator}
\end{equation}
Here $B$ and $\mathcal{J}$ correspond to a sink and a source operator, respectively.
The interaction energy is defined by $\Delta W_n = W_n - 2m_B$ with $W_n$ being
 the $n$-th energy eigenvalue, while the inelastic threshold energy is defined as
 $\Delta W_{\mathrm{th}} = W_\mathrm{th} - 2m_B$
 \footnote{At $m_\pi= 0.51$ GeV, the closest inelastic channel is either  $\Xi^\ast\Xi$ or $\Omega\Sigma$
in $^5D_0$ channel, both of which give $\Delta W_\mathrm{th} \simeq 0.26-0.31$ GeV 
depending on our lattice volumes.
}.
 Below the
threshold (or equivalently $t \gg  (\Delta W_{\mathrm{th}})^{-1}$),
   the correlation $R(\vec{r},t)$ satisfies the time-dependent wave equation, 
\begin{equation}
  \left[
 \frac{1}{4m_B}   
 \frac{\partial^2}{\partial t^2}
 - \frac{\partial}{\partial t} - H_0
  \right] R(\vec{r},t)
  =\int d\vec{r}^{\ \prime}
  U(\vec{r}, \vec{r}^{\ \prime})
  R(\vec{r}^{\ \prime}, t).
\end{equation}
Making the velocity expansion for the non-local kernel $U(\vec{r},\vec{r}^{\ \prime})$,
the leading order (LO) potential becomes
\begin{equation}
  V(\vec{r}) = \frac{1}{4m_B}\frac{(\partial/\partial t)^2 R(\vec{r},t)}{R(\vec{r},t)}
  - \frac{(\partial/\partial t)R(\vec{r},t)}{R(\vec{r},t)}
  - \frac{H_0 R(\vec{r},t)}{R(\vec{r},t)}.
  \label{eq:potential}
\end{equation}
Unlike the case of the L\"{u}scher's method, the time-dependent HAL QCD method does not
require  the ground state saturation of the two-baryon system  as long as  
the contaminations from inelastic states in Eq.~(\ref{R_correlator}) are suppressed.
Also, the interaction kernel (the potential) is spatially localized
and is insensitive to the lattice size. This enables us to calculate  the observables in 
the infinite volume rather easily on the basis of  $ V(\vec{r})$.

\vspace{-1ex}
\subsection{Source operator dependence}
\begin{figure}
  \centering
  \includegraphics[width=0.47\textwidth,clip]{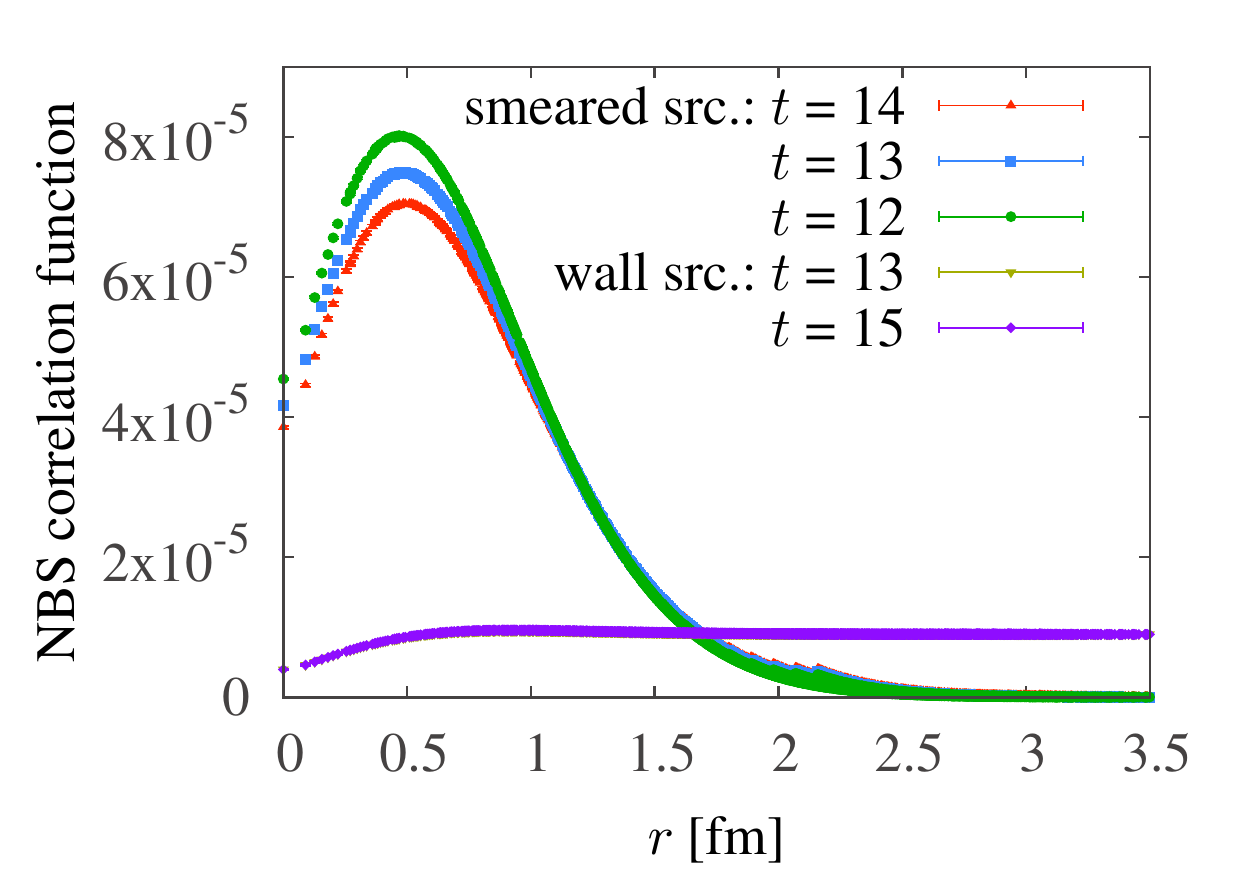}
  \caption{
    \label{fig:wave_function}
    The NBS correlation functions for both smeared and wall sources
    on a $48^3 \times 48$ lattice.
    The smeared source result is localized and  shows strong $t$-dependence,
    while the wall source result is delocalized and almost $t$-independent.
  }
\end{figure}

\vspace{-1ex}
Let us  study the source operator dependence of the potential $V(\vec{r})$ in the HAL QCD method.
In Fig.~\ref{fig:wave_function}, $R(\vec{r},t)$
 with the wall source and the smeared source are shown for $t=12-15$ on a
 $48^3 \times 48$ lattice.
The NBS correlation functions for the smeared source  
are spatially localized and have a visible $t$ dependence.
On the other hand,  the NBS correlation functions
 for the wall source are spatially extended and are insensitive to $t$.

\vspace{-1em}

\begin{figure}
  \centering
  \includegraphics[width=0.47\textwidth,clip]{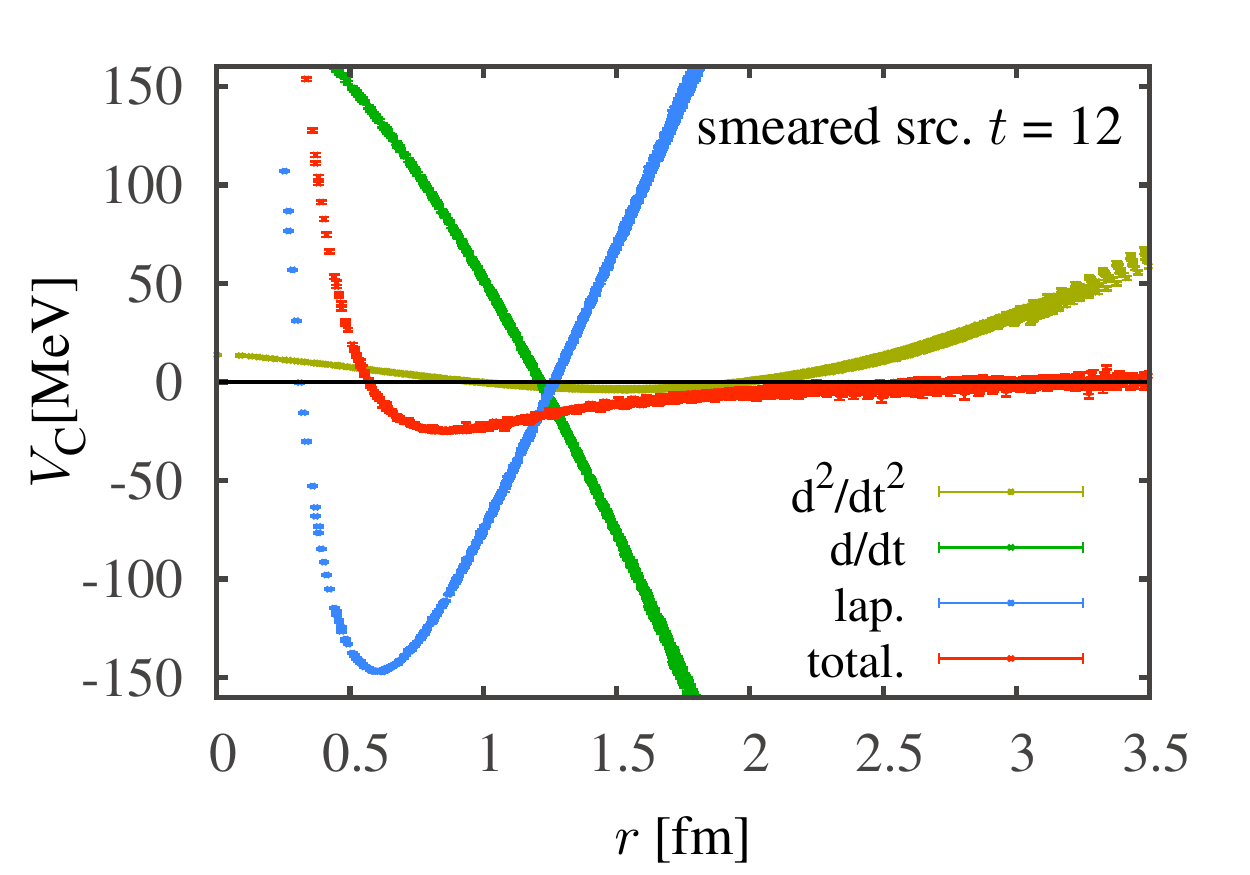}
  \includegraphics[width=0.47\textwidth,clip]{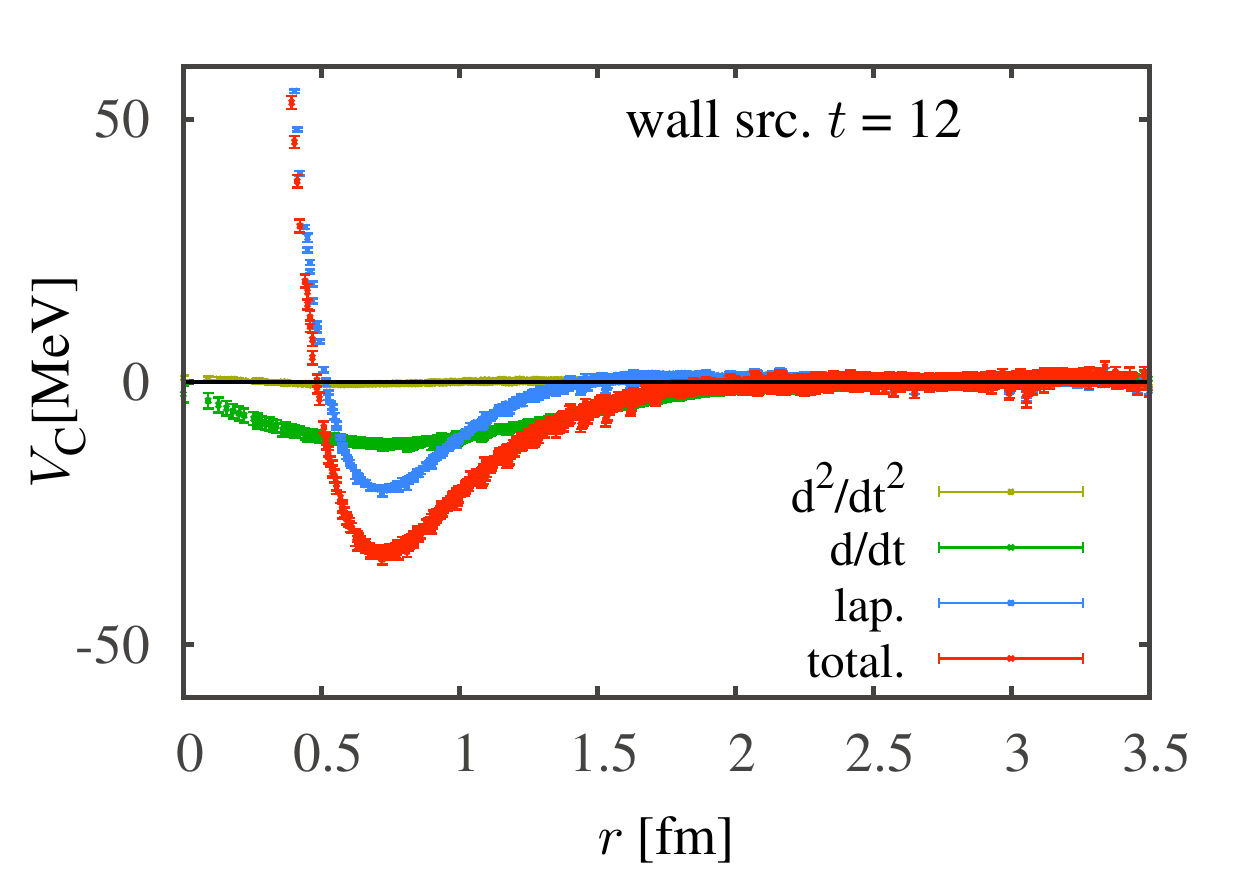}
  \caption{
    \label{fig:potential_breakup}
    Breakdown of the central potential $V_c(r)$ in the $\Xi\Xi(^1\mathrm{S}_0)$ channel
    on a $48^3 \times 48$ lattice.
    (left) The result from the smeared source at $t = 12$.
    (right) The result from the wall source at $t = 12$.
  }
\end{figure}

\vspace{-0.5em}
Although the NBS correlation functions between different sources differ considerably,
the potentials obtained from $R(\vec{r},t)$ tend to be the same \cite{HALQCD:2012aa}.
Shown in Figure~\ref{fig:potential_breakup} is the central potential
 in the  $\Xi\Xi(^1\mathrm{S}_0)$ channel reconstructed from the NBS correlation function at $t=12$
 for the smeared source (left) and for the wall source (right), together with
  its breakdown to each contribution in the right hand side of  Eq.~(\ref{eq:potential}).
For the smeared source, after  sizable  cancellations among different terms, the net result reaches
 to the red symbols.
For the wall source, the cancellation is significantly milder though not negligible.

\vspace{-1.0em}

\begin{figure}
  \centering
  \includegraphics[width=0.47\textwidth,clip]{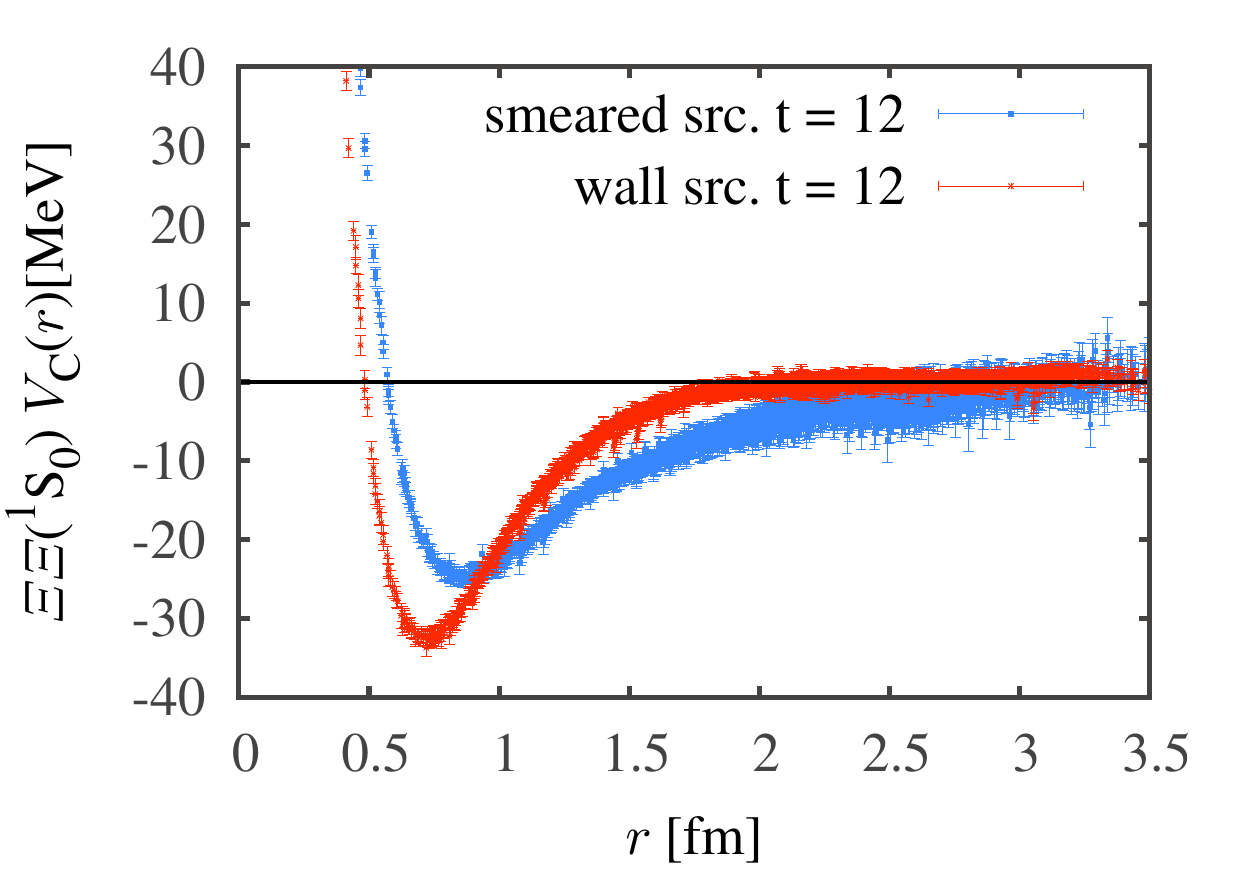}
  \includegraphics[width=0.47\textwidth,clip]{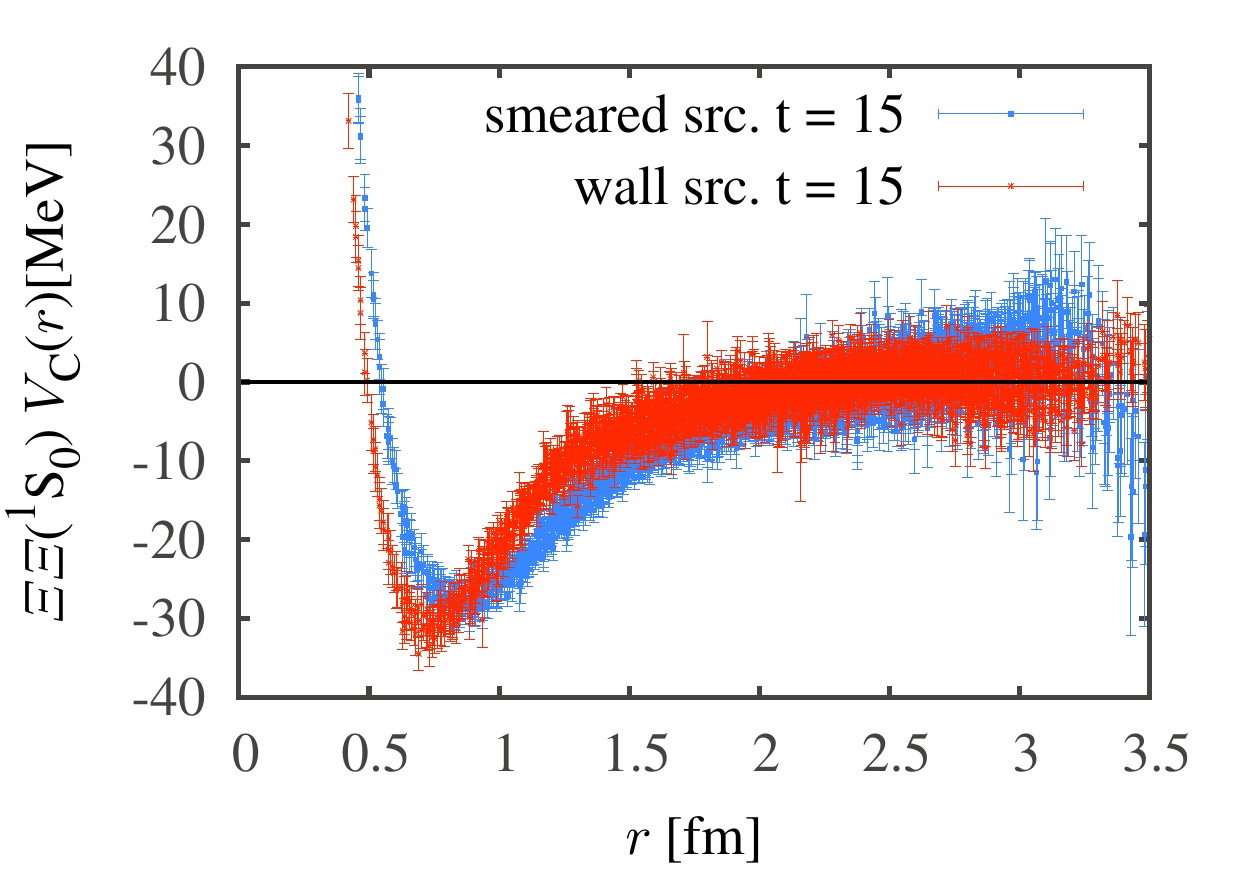}
  \caption{
    \label{fig:potential_comp}
    Comparisons of $\Xi\Xi(^1\mathrm{S}_0)$ central potential 
    by using smeared and wall sources at $t = 12$ and $15$.
  }
\end{figure}

\vspace{-0.5em}
To see the consistency of the potentials from two different sources,
 we show $V_C(r)$ at $t=12$ in Fig.~\ref{fig:potential_comp} (left)
 and that at $t=15$ in Fig.~\ref{fig:potential_comp} (right).
 One finds that the potential of the wall  source is insensitive to the 
 change of $t$ from 12 to 15 within statistical errors. Furthermore, 
 as $t$ increases, the potential  obtained from
 the smeared source approaches to that of the wall source.
   This tendency is also observed in other volumes.
   
\vspace{-2ex}
\subsection{Energy shift from the HAL QCD method}
\begin{figure}
  \centering
  \includegraphics[width=0.47\textwidth,clip]{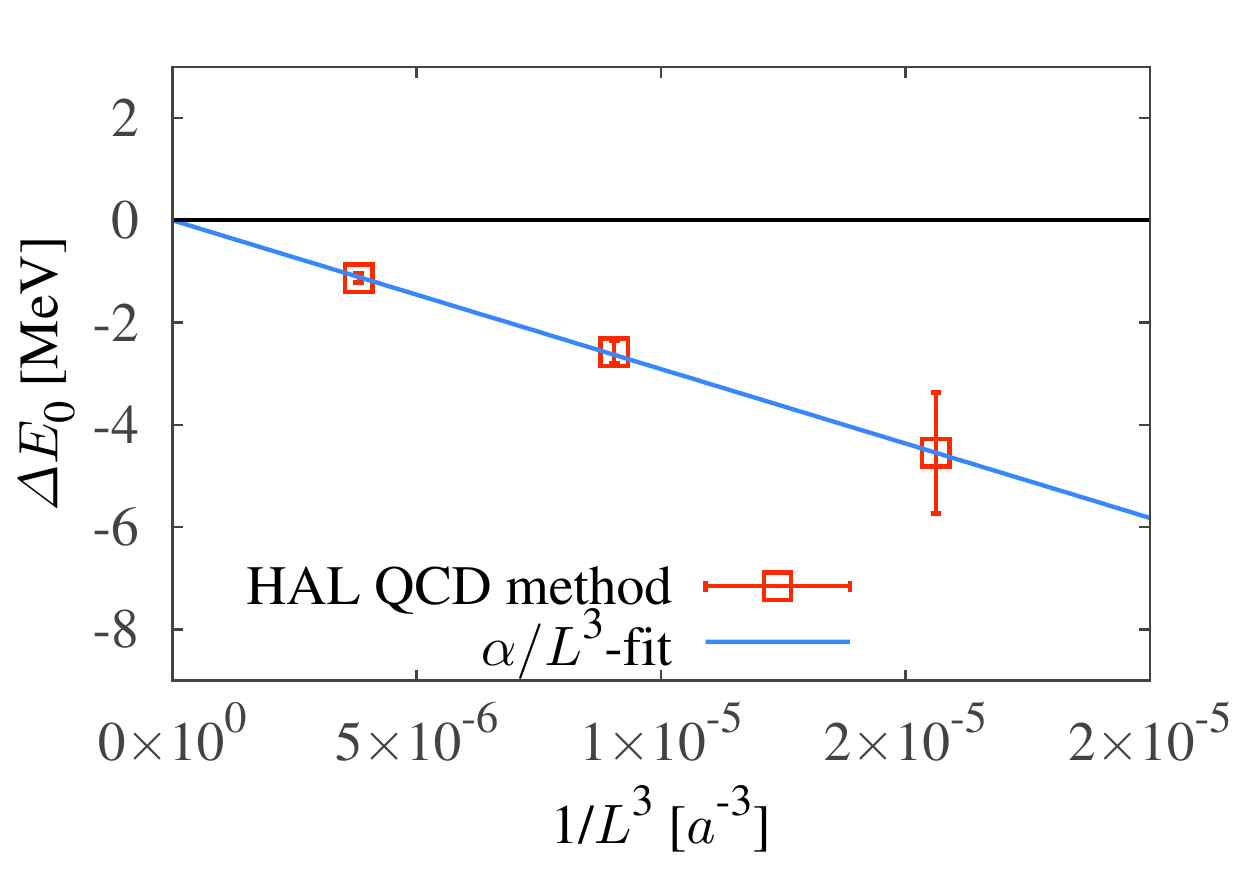}
  \includegraphics[width=0.47\textwidth,clip]{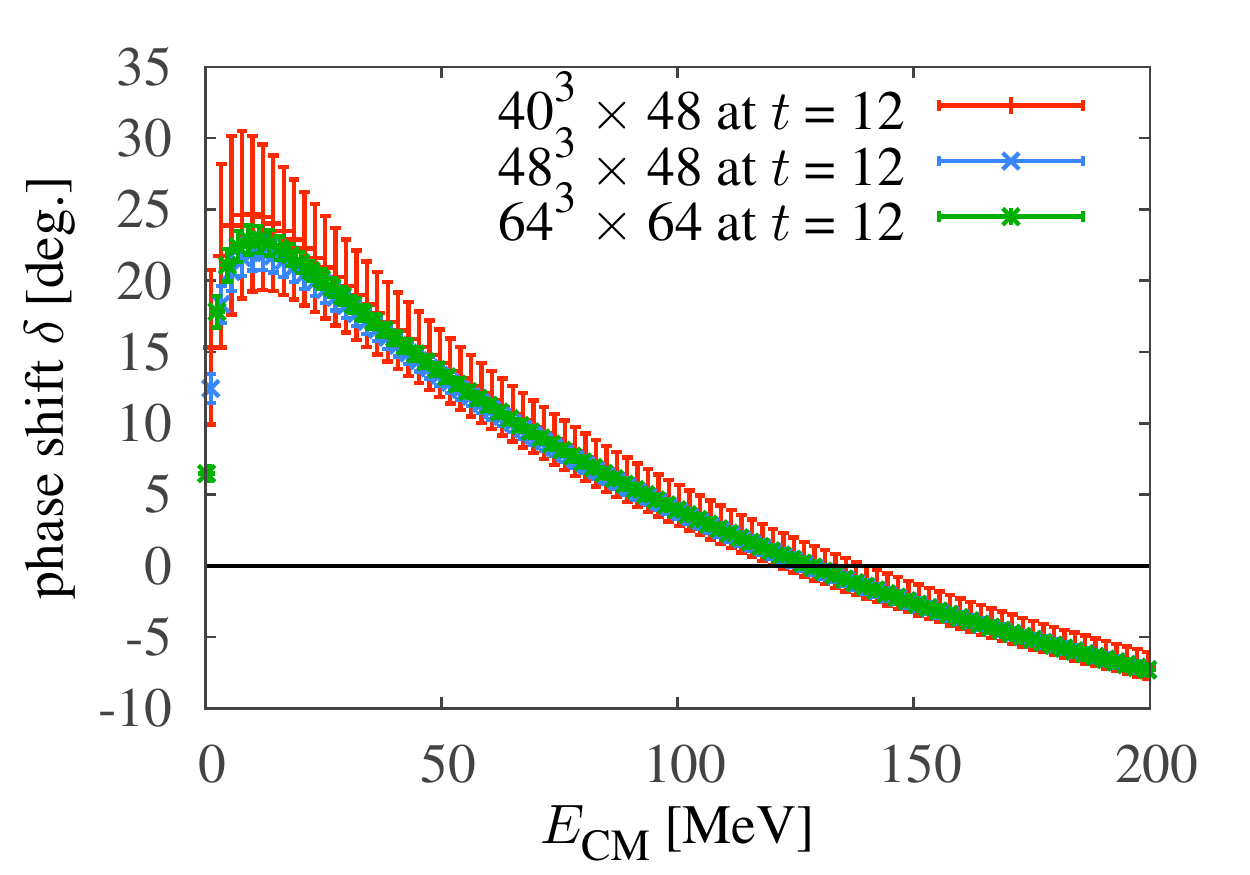}
  \caption{
    \label{fig:vol_dep}
    (left) Volume dependence of the energy eigenvalues of the  ground state. 
    (right) The phase shift of $\Xi\Xi(^1\mathrm{S}_0)$ from the wall source potential, fitted
    by the (two Gaussian + (Yukawa)$^2$) form~\cite{Yamada:2015cra}. 
  }
\end{figure}

\vspace{-1ex}
The source independence of the potential at $t=15$ in
 the previous subsection gives us some confidence on its reliability
 in contrast to $\Delta E_\mathrm{eff}(t)$. 
We can even estimate the possible  energy shift $\Delta E(L)$ in finite $L$
 by using $V(\vec{r})$  \cite{Charron:2013paa}.
 Taking the potential at $t=12$ obtained by the wall source,
 we calculate eigenvalues of $H=H_0+V$ on the lattices with the spatial sizes, $L^3=40^3, 48^3$ and $64^3$.
Resultant energy eigenvalues of the ground and first excited states are summarized in Table~\ref{tab:energy_shift}, and
the lowest eigenvalues $\Delta E_0(L)$ 
are  plotted as a function of $1/L^3$ in Fig.~\ref{fig:vol_dep} (left).
We find that $\Delta E(L)$ behaves linearly in $1/L^3$
and $\Delta E_0(L\rightarrow \infty) \rightarrow 0$.
  In Fig.~\ref{fig:vol_dep} (right), we also show the phase shift 
  from the wall source potential at $t = 12$, fitted by the (two Gaussians + (Yukawa)$^2$)~\cite{Yamada:2015cra}.
Both analyses indicate that the $\Xi\Xi(^1\mathrm{S}_0)$  channel at $m_{\pi}=0.51$ GeV 
has only scattering states in the infinite volume.

\vspace{-1ex}
\begin{table}[h]
  \centering
  \begin{tabular}{cll}
    \hline \hline
    volume & $\Delta E_0$ [MeV] & $\Delta E_1$ [MeV] \\
    \hline
    $40^3$ & $- 4.55 (1.18)$ & $75.63(1.31)$ \\
    $48^3$ & $- 2.58 (22)$ & $52.87(33)$ \\
    $64^3$ & $- 1.13 (9)$ & $28.71(9)$ \\
    \hline \hline
  \end{tabular}
  \caption{The volume dependence of the energy eigenvalues of $H=H_0+V$
   with the potential at $t=12$ with  the  wall source.
  \label{tab:energy_shift}
  }
\end{table}

\vspace{-1em}

\vspace{-5ex}
\section{Summary}
\vspace{-1.5ex}
In this report, we have examined the $\Xi\Xi(^1\mathrm{S}_0)$ interaction 
in 2+1 flavor QCD by the L\"{u}scher's finite volume method and the time-dependent  HAL QCD method.
We have two main conclusions. First of all,
the effective energy shift  $\Delta E_\mathrm{eff}(t)$ 
 shows plateau for $t=12-16$ both  for the wall source and the smeared source, but
  the magnitudes of $\Delta E_\mathrm{eff}(t)$ do not agree  with each other.
  This implies that either one of the  plateaux (or both) is fake. 
  Taking a specific source and extracting the energy shift at relatively small $t$ are quite
  dangerous even if one finds a plateau-like structure. 
   This is because cancellations among scattering states may make a fake plateau in $\Delta E_\mathrm{eff}(t)$ at small $t$.
  A more sophisticated method such as the variational approach discussed in \cite{Charron:2013paa}
  would be necessarily for the reliable extraction of $\Delta E(L)$.
Secondly,    the time-dependent HAL QCD method gives a potential  which is
  insensitive to the choice of the interpolating source operators.
 A tendency toward the source independence can be seen explicitly by
 increasing  $t$.  Using the potential obtained by the HAL QCD method, one can evaluate 
  the energy eigenvalues in finite lattice boxes. The results indicate that
   there is no bound state in the $\Xi\Xi(^1\mathrm{S}_0)$ channel at $m_{\pi}=0.51$ GeV.  
   
   \vspace{-2.7ex}
\section*{Acknowledgements}
\vspace{-1.7ex}
  We thank the authors of \cite{Yamazaki:2012hi} for providing
  the gauge configurations
  and
  the detailed account on the smeared source used in \cite{Yamazaki:2012hi}.
  We also thank authors and maintainers of CPS++~\cite{CPS}, Bridge++~\cite{bridge} and cuLGT~\cite{Schrock:2012fj}
  used in this study.
  The lattice QCD calculations have been performed on Blue Gene/Q at KEK (Nos. 12/13-19, 13/14-22, 14/15-21)
  and HA-PACS at University of Tsukuba (Nos. 13a-23, 14a-20).
    This work is supported in part by the Grant-in-Aid of the Japanese
  Ministry of Education (No. 25287046),
  and the SPIRE (Strategic Program for Innovative REsearch) Field 5 project.

\end{document}